\documentstyle[aps, psfig]{revtex}

\begin{document}

\title{Atom Optical Elements for Bose-Einstein Condensates}
\author{S. \ Choi \ and \ K.\ Burnett}
\address{Clarendon Laboratory, 
Department of Physics, University of Oxford, Parks Road, 
\mbox{Oxford OX1 3PU, United Kingdom.}}
\maketitle

\vspace{6mm}
\begin{abstract}
A simple model for atom optical elements for Bose condensate of trapped, dilute alkali atoms is proposed and numerical simulations are presented to illustrate its characteristics. We demonstrate ways of focusing and splitting the condensate by modifying experimentally adjustable parameters.  We show that there are at least two ways of implementing atom optical elements: one may modulate the interatomic scattering length in space, or alternatively,  use a sinusoidal, externally applied potential. 
\end{abstract}


\section{Introduction}
Since the first experimental realisation of Bose-Einstein Condensation (BEC) with dilute, trapped alkali gases in 1995\cite{BECRb,BECNa,BECLi}, rapid progress has been made in studying their physical properties.   One of the exciting prospects for BEC from the outset has been the possibility of realising the matter equivalent of optical laser, i.e.  the atom laser\cite{HolBurGar96,WisCol95}. 
The recent demonstration by the MIT group of an output coupler for condensates\cite{MewAndKur97} along with the interference of two independent condensates\cite{AndTowMie97} has brought the realisation of the atom laser one step closer. In fact, one could argue that we already have a rudimentary atom laser and that we should be looking for effective ways to control the output beam of such sources.  By control we mean focusing and beam splitting of flowing condensates, in analogy to conventional optics using light. In the last few years, there has been tremendous progress made in the field of atom optics, both linear and non-linear, in which the motion of atoms is controlled via light forces\cite{AdaSigMly94,DroStuSch97,LenMeyWri93,Zha93,ZhaWalSan94,ZhaWal94}. Direct application of these techniques using light forces will not always be possible due to a variety of factors. These include the possible heating of the condensate by the laser field which may destroy the condensation. Ideally one wants to apply light fields to transfer momentum to the atoms in a controlled manner while keeping the atom assembly condensed. This is, of course, a particularly severe task for ultracold BEC's. 

In this paper we apply the well-known idea of optically generated potentials used in atom optics in the context of Gross-Pitevskii Equation (GPE) to model how the atom optical elements produce focusing and splitting of a condensate. We should emphasise that we shall be demonstrating the effect of such arrangements rather than providing detailed descriptions of how such a configuration would be constructed in practice. We shall, however, discuss some of what we see as the main limitations of such schemes. In section II we shall specify our theoretical model and in section III the results of our simulation are presented. Finally, these results are discussed in view of available experimental parameters, in section IV. 

\section{Model}

\subsection{The NLSE or Gross-Pitaevskii Equation}
The mean-field theory for an assembly of condensed atoms in an external field may be obtained in two ways: in the first, we start with the non-linear atom optics point of view of many-body wave function and then find an appropriate limit for Bose condensation; in the second, we use a field theory approach of Ginzburg and Landau\cite{GPE} which is the standard approach in much of the literature on Bose condensates. The mean field theory obtained from these two approaches is identical, and we shall briefly outline only the second route below. For a detailed discussion of formalism of nonlinear atom optics, readers are referred to papers by Lenz {\it et al.}\cite{LenMeyWri93} and Zhang {\it et al.}\cite{Zha93,ZhaWalSan94,ZhaWal94}.  

The Hamiltonian for a system of interacting atoms in a trap is given by:
\begin{equation}
\hat{H} = \int \hat{\Psi}^\dagger({\bf r},t)\hat{H_0}\hat{\Psi}({\bf r},t)d^3{\bf r} + \frac12 \int \int \hat{\Psi}^\dagger({\bf r'},t)\hat{\Psi}^\dagger({\bf r},t)\hat{V}({\bf r},{\bf r'})\hat{\Psi}({\bf r'},t)\hat{\Psi}({\bf r},t),
\label{Hamiltonian}
\end{equation}
where $\hat{H_0}$ is the single particle Hamiltonian
\begin{equation}
\hat{H_0} = - \frac{\hbar^2}{2m}\nabla^{2}_{r} + V_{\rm trap}({\bf r}),
\label{SPHamiltonian}
\end{equation}
and the field operator $\hat{\Psi}$ obeys the standard boson commutation relation
\begin{equation}
[\hat{\Psi}({\bf r}), \hat{\Psi}^{\dagger}({\bf r'})] = \delta({\bf r} - {\bf r'}).
\end{equation}
We shall assume, given the fact that we are dealing with a dilute gas, that the interatomic potential can be approximated by that of contact interaction:
\begin{equation}
\hat{V}({\bf r}, {\bf r'}) = {\cal V}\delta({\bf r}-{\bf r'}), \label{V}
\end{equation}
Additionally, we shall suppose that, 
\begin{equation}
{\cal V} \equiv {\cal V}_{\rm intrinsic} + {\cal V}_{\rm induced}.
\end{equation}
That is to say, the strength of the overall interatomic interaction may be viewed as the sum of that due to intrinsic interactions and the modification due to an external influence such as that of the laser, magnetic field or rf-induced interaction.  Such externally induced modification may occur, for example, when pairs of atoms, driven by the external field,  have the ground state mixed with an electronically excited quasimolecular state. This quasimolecular state can have much stronger interaction, modifying the scattering amplitude in the process.  Description of the interaction strength is facilitated by introducing the scattering length $a$, which is proportional to  ${\cal V}$.  The nonlinear Schr\"{o}dinger equation for the condensate wave function, $\psi$, is then given by: 
\begin{equation}
i \hbar \frac{\partial \psi}{\partial t} = \left [ -\frac{\hbar^2}{2m}\nabla^{2}_{r} + V_{\rm trap}({\bf r}) + C_{\rm n} |\psi|^2 \right ] \psi, \label{NSE}
\end{equation}
where 
\begin{eqnarray}
C_{\rm n} & = & \frac{4 \pi N \hbar^{2}}{m}(a_{\rm intrinsic} + a_{\rm induced})\\ 
          & = & \frac{4 \pi N \hbar^{2}a}{m},
\end{eqnarray}
with $N$ being the number of particles in the sample, and $m$ the mass of a single particle. $a_{\rm intrinsic}$ denotes the scattering length in the absence of an external field and is an experimentally measurable quantity; $a_{\rm induced}$ is that induced by an external electromagnetic field for which there are various theoretical predictions.  It is predicted, for example, that the overall scattering length, $a$, may be varied by the application of a magnetic field for $F=  3$, $m_{F} = -3$ hyperfine state in Cs\cite{TieMoeVer92}. A blue detuned ``optical shielding'' type laser of variable intensity\cite{SuoHolBur95,NapWeiJul97} is also believed to be a good way to achieve control over the effective scattering length. Experimental measurement of an $a_{\rm induced}$, however, has not been performed to date. We note here that, in theory, the non-linear coefficient, $C_{\rm n}$ may be made position dependent, since the induced scattering length, $a_{\rm induced}$, may be varied in space with an application of a laser field with a suitable variation in intensity. 

The nonlinear Schr\"{o}dinger equation or the Gross-Pitaevskii Equation,  Eq. (\ref{NSE}),  describes the dynamics of a single atom moving in the presence of mean field induced by other atoms in the system, in addition to the effects of the confining trap potential.   The GPE has been found, over the last couple of years, to provide very good description of condensate behaviour over a range of temperatures from near $T=0$~K to a sizeable fraction of $T/T_{c}$ where $T_{c}$ is the critical temperature of phase transition\cite{EdwRupBur96,JinEnsMat96,Str96,MewAndKur96}.

\subsection{Modified GPE}

We intend to investigate two types of possible experimental arrangement. As the first possibility, we consider a typical atom optical configuration as shown schematically in Fig 1: Firstly the condensate is prepared in a trap and released so that it freely falls under gravity. Such a situation is described by turning the trap off i.e. setting $V_{\rm trap} = 0$. The cloud of atoms will undergo ballistic expansion. The condensate is then passed through an interaction region in which it is subject to an interaction potential, before it again falls freely in space onto a detector or a surface.

We look at two types of interaction with this set up: spatial modulation of the scattering length in the interaction region, so that the corresponding evolution is given by: 
\begin{equation}
i \hbar \frac{\partial \psi}{\partial t} = \left [ -\frac{\hbar^2}{2m}\nabla^{2}_{r} + {\cal C}(\cos {\bf k \cdot r} + {\cal D}) |\psi|^2 \right ] \psi,  \label{Cvary}
\end{equation}
and use of sinusoidal potential in the interaction region as described by: 
\begin{equation}
i \hbar \frac{\partial \psi}{\partial t} = \left [ -\frac{\hbar^2}{2m}\nabla^{2}_{r} +{\cal A}\cos {\bf k \cdot r} + C_{\rm n} |\psi|^2 \right ] \psi,  \label{Vvary}
\end{equation}
where ${\cal C}$ and ${\cal A}$ are the amplitudes and $k$ the wave number which can be adjusted. We present the results for the case when ${\cal D} = 0$, and also the case when the sinusoidal variation is limited to positive scattering lengths, by a suitable adjustment of ${\cal D}$. 

Using either method, the atomic motion may be modified in a controllable manner.   The two methods are, however, qualitatively different since the first involving $C_{\rm n}$ is of an intrinsic nature, which arises from interaction of atoms in a many body system while the second involves imposing an external potential. We shall see that unlike conventional atom optics, we now have two avenues to impose potential gradient, which may or may not be applied simultaneously and opens up the possibility of manipulating the relative amplitudes and phases of these two potentials. In this way, it seems possible to generate an effective overall potential gradient, which may even vary over time, to tailor our atom optical component.

As the second possible experimental arrangement, we look at the effect of spatially varying the interatomic scattering length in the presence of the confining trap. The condensate is not dropped, and therefore we can study the case in which one has both types of potentials present simultaneously.
This is modelled by an equation with a minor modification to the standard GPE: 
\begin{equation}
i \hbar \frac{\partial \psi}{\partial t} = \left [ -\frac{\hbar^2}{2m}\nabla^{2}_{r} + V_{\rm trap}({\bf r}) + f({\bf r})|\psi|^2 \right ] \psi.  \label{VC}
\end{equation}
We consider the case of $f({\bf r}) = {\cal G}{\bf r}^2$ as well as $f({\bf r}) = {\cal C}(\cos {\bf k \cdot r} + {\cal D})$, where ${\cal G}$, ${\cal C}$, ${\cal D}$ are constants, in order to test general features of such an arrangement.   

The condensate formed recently at MIT using the cloverleaf trap configuration was reported to have an aspect ratio of 20\cite{MewAndvan96}. This means that a 1 dimensional GPE may be expected to provide an adequate qualitative description of the evolution of a cross section of such condensate. The spatial modulation  of $C_{\rm n}$ discussed above would, in this case, be along the long axis.  To be more precise, one may allow some Gaussian spreading of $C_{\rm n}$ in the radial direction of the cigar-shaped condensate. i.e. $C_{\rm n}({\bf r})_{\rm radial}$ of the form
\begin{equation}
C_{\rm n}({\bf r})_{\rm radial} \propto  e^{-\kappa {\bf r}_{\rm radial}^2}.
\end{equation}
We assume the coupling between the $C_{\rm n}({\bf r})_{\rm axial}$ and  $C_{\rm n}({\bf r})_{\rm radial}$ to be weak.  The 1D GPE with modified linear and non-linear potentials was integrated using a 4th order Runge-Kutta integration routine over various configuration times in order to find the optimal condition for focusing and splitting.   In the following we assume that all our atoms originally have positive scattering lengths. 

\section{Results}

\subsection{Results for the atom optical configuration}
In order to simulate the experiment of type shown in Fig. 1, we need to specify how long we let the condensate to free fall, interact with the chosen field and then fall freely again onto a detector or a substrate.  
It was found from simulations that, once the initial free evolution time is chosen, the final spatial distribution of the condensate that we see at the detector depends, as would be expected, quite sensitively on the duration of interaction and the subsequent free fall. In our simulations, we chose a reasonable initial free fall time and then selected the interaction and the subsequent free fall times so as to give optimum probability distribution in terms of sharpness and regularity of the peaks on the final detector.     

\subsubsection{Spatial modulation $C_{\rm n}$ in the interaction region}
In conventional atom optics, parabolic optical potentials are used to provide a focussing effect. We have, therefore, tried a parabolic modulation of $C_{\rm n}$ in the interaction region. 
It was found that, in order to see any noticeable focusing effect by such an arrangement, we need a very steep position dependence, such that around the tail region of the condensate, $C_{\rm n}$ of the order of around 20 times the original value is needed. A theoretical treatment by Fedichev {\it et al.}\cite{FedKagShl97} indicates that although it might be possible to change the scattering length by comparable amounts, the accompanying inelastic scattering rate also undergoes a correspondingly large increase.  Since this is not experimentally favourable, we do not pursue the parabolic variation any further and instead concentrate on the sinusoidal modulation of $C_{\rm n}$ in the interaction region.

We have let our condensate to ballistically expand for $t = 1\pi/\omega$ to model the initial free fall part of Fig. 1.  The axial trap frequency  of the MIT cloverleaf trap was reported to be $\omega = 2\pi \times 18$ Hz which translates $t$ to time of order 27 ms.  Figure 2 shows in solid line the probability distribution immediately after the interaction region, interacting for duration $t = 0.3\pi/\omega$ with modulation amplitude ${\cal C} = C_{\rm n}$, ${\cal D} = 0$  and wave number $k = 2$ in harmonic oscillator units. The dashed line indicates the ballistically expanded condensate just before the interaction. This interaction time, which corresponds to approximately 8 ms in the MIT trap parameters,  gave the best result in terms of sharpness, height and regularity of the peaks.  The shape of the spatial modulation is shown as the dotted line. We see that the number of peaks and valleys match the number of spikes in the condensate as to be expected from conventional potential gradient idea. Indeed, it was found that in an identical simulation with $k =1$ i.e. half the wave number, we get 4 prominent spikes in the final distribution instead of 8 as seen here.  It is then reasonable to expect that the sharpness of these peaks, as indicated by the ratio of their height to widths would increase if the amplitude of the potential was increased; the results agree with this prediction.  The maximum  peak height of 0.094 for Fig. 2 was increased to 0.162 as a result of doubling the amplitude of modulation while keeping the general shape. 

After the interaction region, the condensate was let to evolve freely. Figure 3 shows the final probability distribution after the distribution shown in Fig. 2  was freely evolved for duration $t = 0.3 \pi/\omega$.  The time of free evolution was again chosen so as to give the best distribution. The peaks are now broader and have been displaced so that now there is a peak at the origin, as a result of rather complex dynamical motion of various parts of the condensate as it expands.  The dotted lines represent the shape of the effective potential that was originally in place in the interaction region. It was found that a regular probability distribution is again restored after $t = 0.8\pi/\omega$, with shorter and broader peaks. 

The case of a sinusoidal variation of the scattering length which remains positive was also investigated. This was simply accomplished by translating the potential vertically so that the minimum value of $C_{\rm n}$ is now 0 instead of $-C_{\rm n}$ as in the previous case. The final result is shown in Fig. 4. The solid line is the case when $C_{\rm n}$ varies from $0$ to $C_{\rm n}$. The dot-dashed line indicate the case when 
$C_{\rm n}$ varies from 0 to $2C_{\rm n}$.  We see that there is no qualitative difference from the previous case in which negative scattering length was allowed; the amplitude of modulation seems to be the determining factor rather than the sign.

The effect of longer initial free evolution time than $1\pi/\omega$ was to give shorter peaks, but with the same general shape and width. This can be explained from the fact that after a longer free fall, the condensate has spread out more and its peak height is consequently shorter when it interacts with fields in the interaction region. 

\subsubsection{Using a sinusoidal potential in the interaction region}
In a direct analogue to the case of spatial modulation of $C_{\rm n}$, we try and simulate the same experiment of Fig. 1, but here we employ a sinusoidal potential in the interaction region while leaving $C_{\rm n}$ spatially invariant i.e.  Eq. (\ref{Vvary}).     The initial free fall time was again chosen to be $1\pi/\omega$. 

The probability distribution immediately after the interaction region of Fig. 1 for a sinusoidal potential is shown as solid line in Fig. 5. In this case, Eq. (\ref{Vvary}) was integrated with ${\cal A} = 1$ and $k = 2$ over interaction time $t = 0.2 \pi/\omega$. As before, the shape of the sinusoidal potential is shown in dotted line, and the dashed line indicates the ballistically expanded condensate immediately before the interaction.  When the amplitude of modulation was doubled, the maximum  peak height of 0.073 shown in Fig. 5 was found to increase to  0.102 while keeping the same general shape.  Again, when an initial free evolution time longer than $1\pi/\omega$ was considered, shorter peaks were produced, giving smaller peak height-to-width ratio, as discussed previously.

The final probability distribution after a subsequent free evolution of duration $t = 0.3 \pi/\omega$ is shown as solid line in Fig. 6.  It was again found that a regular probability distribution is restored after $t = 0.9\pi/\omega$, with shorter and broader peaks. 
When one uses ${\cal A} = 2$ in the sinusoidal potential, the corresponding probability distribution is as given by the dot-dash line in Fig. 6. For this case, free evolution time after the interaction of $0.4 \pi/\omega$ was used. We see here that the effect of  modifying the confining trap potential is qualitatively the same as that of modulating $C_{\rm n}$ and that their effect in both cases may be understood by simply using the potential gradient idea, as in conventional atom optics.

We note also that parabolic potential over the interaction region is another possible configuration for focusing. It was found, however, that although focusing effect does arise in the sense of counteracting the ballistic expansion, its effectiveness was rather small and broadening took place on a very short time scale after leaving the interaction region. This is possibly because by this stage, repulsive interactions dominate in a dynamically expanding condensate, such that the original trap potential is not sufficient to counteract the expansion. Again a very steep parabolic potential was needed to observe an efficient focusing effect.

\subsection{Spatial modulation of $C_{\rm n}$ in the presence of a trap}
In this subsection, we look at the possibility of imposing both the spatial modulation of $C_{\rm n}$ and a confining trap. We look at the evolution such as that described by Eq. (\ref{VC}), which, in addition to the atom optical configuration discussed in the previous subsection, is  another possible experimental arrangement. Since, in this case, the confining trap prevents ballistic expansion, a parabolic variation in $C_{\rm n}$
with more realistic values of the scattering length could be studied. 

Figure 7 displays the probability distribution of the original trapped condensate in solid line while the dot-dash line displays the distribution with a parabolic variation on the non linear coefficient in Eq. (\ref{VC}). This is the result after time $t= 0.5 \pi/\omega$ where $\omega$ is the trap frequency.  In this simulation we assumed that $C_{\rm n}$ varies as $C_{\rm n}x^{2}/400$ in harmonic oscillator units so that we have  $C_{\rm n}(x) = 0$ at the origin and $C_{\rm n}$ at $x=\pm 20$. We see here a focusing effect within the trap. 

The height, and consequently the width, of the peak was found to vary over time in an oscillatory fashion. Figure 8 shows the evolution of the maximum height of the peak over time in the presence of this parabolic potential. The behaviour can be qualitatively understood from the fact that the condensate on the ``wings'' of the gaussian-like initial distribution `feels' a push towards the center, resulting in the build up of the peak height; the peak height then starts to decrease after a maximum as the condensate pass through each other and results in broadening of the base of the distribution. 
A slowly-varying beating type envelope appears on the oscillatory curve when a different parabolic distribution is used; for instance when the $C_{\rm n}$ varies as $C_{\rm n}x^{2}/100$ rather than $C_{\rm n}x^{2}/400$.
This slowly-varying beat pattern is caused by effective phase difference between the potentials acting on the condensate. In this example, the potentials act as two  distinct ``springs'' which are out of phase with one another in time; this effect becomes more pronounced when the tightness of the parabolic potential due to $C_{\rm n}$ is changed.

The resulting probability distribution when a sinusoidal rather than a parabolic variation of scattering length is imposed is shown in Fig. 9. We see that an action similar to beam splitting occurs with two prominent peaks. The time of interaction shown is $t = 0.4\pi/\omega$. Again there is a time dependent behaviour as to be expected.

The somewhat unexpected shape of Fig. 9 can be explained by superposing the two independent potentials;  Fig. 10 shows the shape of the overall potential when the parabolic trapping potential and the sinusoidal one modulating the non-linear constant is superposed. (This is a plot of $x$ vs. $x^2/4 + C_{\rm n}\cos x$.) It is clear from the shape that the most likely probability distribution is that of two peaks, as any further peaks would require additional energy. Again, the time-dependent variation of the probability distribution can be explained by the fact that, as the condensate changes its shape over time, both the trapping potential and the sinusoidal potential from the nonlinearity exert time-varying forces.  It is noted that in Fig. 9, we used $k= 1$ modulation, and the separation between the two peaks was reduced for higher $k$, as to be expected.  When the trap was turned off, it showed a lot more peaks than the expected number of around 4 (c.f. Figs. 4-6). This is readily seen to be the effect of the trapping potential; there is an extra force from the sides toward the center, resulting in high number of peaks in the main region $x = -10 \ldots 10$. What is clear from simulations of this subsection is that we can manipulate the shape of an intially stationary condensate within a trap by modulating the non-linearity constant.

\section{Discussion}

In this section we attempt to estimate possible experimental values.  First of all, we note that the interaction times used in simulations of the order of 10~ms are very much within the limits of present technology, since a typical atom lithography set up has the total time taken from a thermal source to substrate of the order  0.67 ms\cite{DroStuSch97}. It is also noted that typical time of flight of the condensate used in recent experiments before imaging is around 40~ms, so the total time taken from source to detector in our numerical simulations matches those of real experiment very closely.  The total distance the condensate falls in our example from source to detector is, then, around 1~cm.  The width of each peak in simulations is around 2 harmonic oscillator units which, for sodium atoms and the cloverleaf trap, corresponds to approximately 7 $\mu$m. 
Possible ways to reduce this width even further would be by using the modulating field of higher value of amplitude to $k$ ratio in the interaction region and also by using a trap of higher frequency than $2\pi \times 18$ Hz. The effect of increasing $k$ alone would mean the peaks become more closely spaced.

The area under each peak was such that the number of atoms in each peak would be around 10\% of total number of atoms.   A very important point is whether there is any limit to the number of atoms in each peak due to interactions between the particles. This aspect was investigated by looking at the probability distributions for the case where we employ sinusoidal potential, i.e. Eq. (\ref{Vvary}), but with various values of interatomic interaction strength characterised by $C_{\rm n}$. It was indeed found that the ``fringe visibility'' defined as  
\begin{equation}
V =\frac{P_{\rm max} - P_{\rm min}}{P_{\rm max} + P_{\rm min}},
\end{equation}
where $P_{\rm max}$ and $P_{\rm min}$ represent the maximum and the minimum of the ``fringe pattern'' decreases quite dramatically when we use a larger value of $C_{\rm n}$.  As an approximate indication, simulation with $C_{\rm n} = 150$, as compared to $C_{\rm n} = 20$ used in Figs 5 and 6 gives  $V= 0.24$ as compared to original $V=0.72$.  The times used were $0.5 \pi/\omega$, $0.2 \pi/\omega$, and $0.3 \pi/\omega$. The initial free evolution time was chosen to be $t = 0.5\pi/\omega$ rather than $1\pi/\omega$  because in this case of large $C_{\rm n}$, the cloud spreads out too much by the time $t = 1\pi/\omega$.  
Noting that $C_{\rm n}$ is also proportional to $N$, the total number of the condensate particles, this brings us to the question of whether production of an {\em intense} beam of atoms in an atom laser is a requirement when atom optical elements of this kind is used. This aspect seems to be an important limiting factor in the design and implentation of any atom optical experiment using bose condensates. One may, in principle, use a higher value of ${\cal A}$ for high values of $C_{\rm n}$ to improve the visibility.  This implies that to be effective,  various parameters have to be adjusted carefully to optimise the set up for each instance. 

Modulation of scattering length and use of sinusoidally varying potential raise the issue of experimental feasibility, as these will certainly involve interaction of an electromagnetic field with ultracold atoms. 
The primary effect of light on cold trapped atoms is to cause loss processes by giving atoms enough energy to escape the trap, through such processes as photon recoil and photoassociation. The rate at which an atom  heats up as a result of photon recoil is given by 
\begin{equation}
\Gamma = \gamma \left ( \frac{\Omega}{\delta} \right )^{2}
\end{equation} 
where $\gamma$ is the natural atomic linewidth, $\Omega$ the Rabi frequency and $\delta$ the detuning. It is assumed here that $|\delta| \gg \Omega$ and $|\delta| \gg \gamma$.   A recent theoretical study by Fedichev {\it et al.} shows that it is, in fact, possible to alter the scattering length by as much as 350\%  for $^7$Li and minimize recoil or photoassociation losses with a selection of frequency detuning and Rabi frequency\cite{FedKagShl97}.   In order to have small recoil losses along with a significant change of $a$ it is necessary that the detuning, $\delta$ be large and negative, and chosen to be not too far from a vibrational resonance with one of the bound $p$ states of the electronically excited quasimolecule.  All our simulations depend on the assumption that the possible heating of the condensate is either negligible, or has typical time scale which is long compared to the interaction times considered here. We also note that generation of optical potentials by Raman transition has recently been proposed as a way to reduce the spontaneous emission rate\cite{HopSav96}. The atom optical elements discussed in this paper may be achieved via magnetic as well as optical fields, and it is conceivable that certain arrangement such as, say, an optical shielding laser combined with a magnetic field-induced potential gradient might provide a solution. More study, however, would be required on exactly how one can implement these atom optical elements. 

In summary, it was found from numerical simulations involving various configurations that focusing and splitting effects can be achieved within the context of the non-linear Schr\"{o}dinger equation with modified potentials. It is pointed out that our simulations just show a possible theoretical model, without inclusion of such effects as the spontaneous emission of atoms as might be the case when the external potential involves an optical dipole force. Various possible improvements may be made in order to accommodate such effects into theory. For example the effect of inelastic collisions may be incorporated by letting $C_{\rm n}$ be complex. However, at present, there is no clear indication of how big such effects will be in practice.  Further studies along the lines of atom optics could involve the study of coherence property of the condensate which have been modified in this way. Other possibilities include finding a way to use the two potentials simultaneously to tailor an atom optical element of an arbitrary specification, and eventually finding a way to prepare vortex states in the condensate through such modification. This will then be of relevance to the study of superfluidity in trapped alkali bose gas, and in this sense, the atom optical elements would be more than just equivalents of conventional optical elements. 

\acknowledgments
This work was supported by UK EPSRC, and the European Union under the TMR Network programme. SC would also like to thank UK CVCP for support.

\vspace{10mm} 
\newpage

\begin{figure}[t] \begin{center}
\centerline{\psfig{height=7cm,file=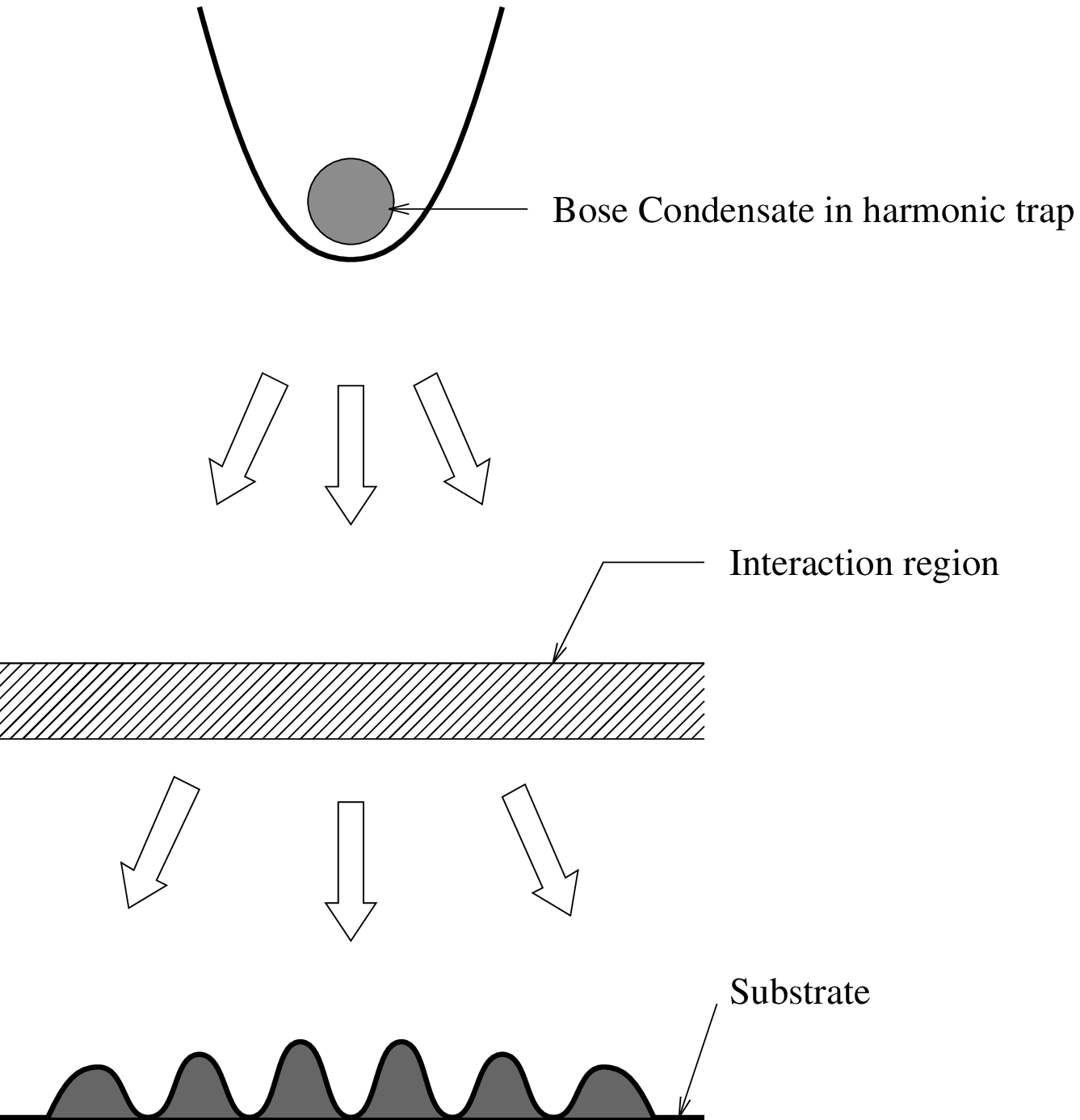}}
\end{center}
\caption{\protect \footnotesize  Schematic diagram of possible experimental configuration. The condensate prepared in a trap is released so that it passes through an interaction region in which it is subject to a modified potential. It then falls freely in space onto the surface of a substrate or a detector.   }
\end{figure}

\begin{figure}[t] \begin{center}
\centerline{\psfig{height=7cm,file=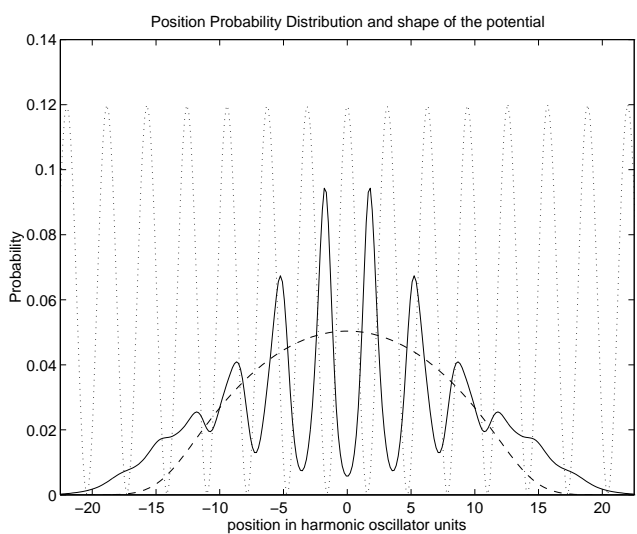}}
\end{center}
\caption{\protect \footnotesize Probability distribution immediately after the interaction region of Fig. 1 for the case of modulating the scattering length, with amplitude ${\cal C} = C_{\rm n}$, ${\cal D} = 0$ and wave number $k = 2$ in harmonic oscillator units (solid line). This is the result after interacting for the duration $t = 0.3\pi/\omega$ in the interaction region. The dotted line indicates the shape of the potential. The dashed line is the shape of the condensate just before the interaction.}
\end{figure}

\begin{figure}[t] \begin{center}
\centerline{\psfig{height=7cm,file=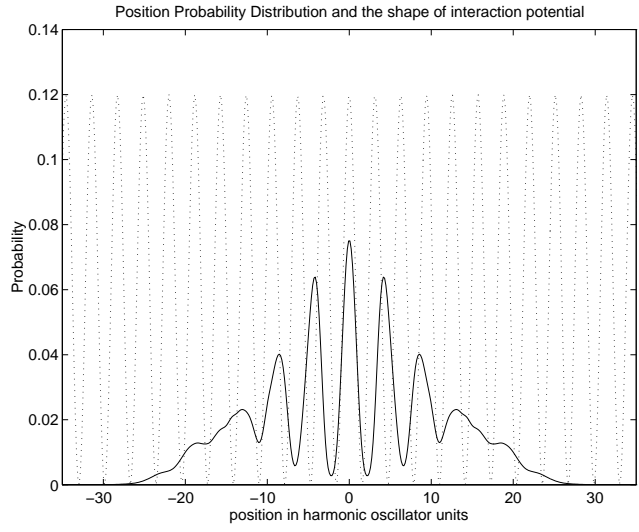}}
\end{center}
\caption{\protect \footnotesize    Final probability distribution after the free evolution of the distribution shown in Fig. 2 for duration $t = 0.3 \pi/\omega$.  The dotted line represents the shape of the potential that was in place in the interaction region.}
\end{figure}

\begin{figure}[t] 
\centerline{\psfig{height=7cm,file=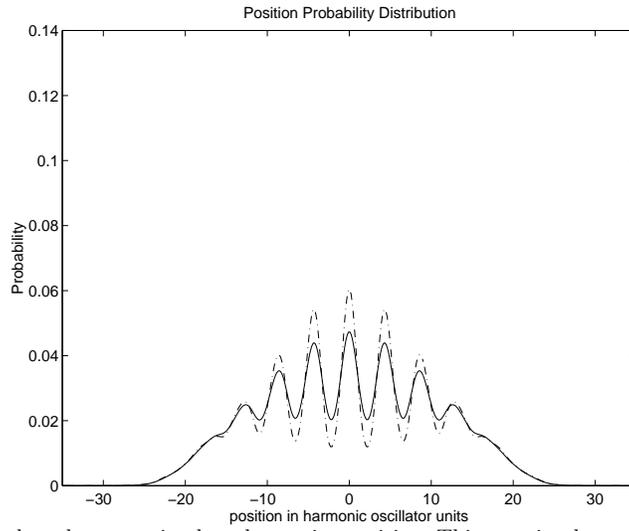}}
\caption{\protect \footnotesize   Probability distribution when the scattering length remains positive. This was simply accomplished by translating the potential vertically (${\cal D} = 1$). Solid line represents the case of varying the potential between 0 and $C_{\rm n}$ while the dot-dash curve corresponds to the potential which varies between 0 and $2C_{\rm n}$.}
\end{figure}

\begin{figure}[t] \begin{center}
\centerline{\psfig{height=7cm,file=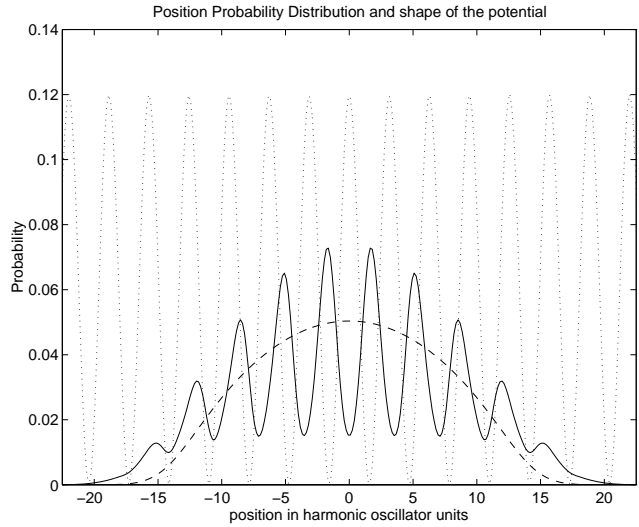}}
\end{center}
\caption{\protect \footnotesize    Probability distribution immediately after the interaction region for a sinusoidal potential. ${\cal A} = 1$ and $k = 2$ with an interaction time $t = 0.2\pi/\omega$ (solid line). The dotted line indicates the shape of the potential, and the dashed curve shows the condensate just before the interaction.}
\end{figure}

\begin{figure}[t] \begin{center}
\centerline{\psfig{height=7cm,file=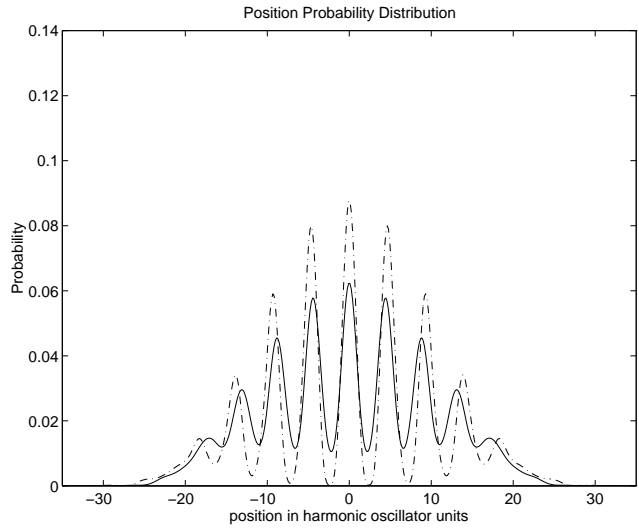}}
\end{center}
\caption{\protect \footnotesize   Final probability distribution after the free evolution of the distribution shown in Fig. 5 $({\cal A} = 1, k = 2)$ for duration $t = 0.3 \pi/\omega$ (solid line).  The dot-dash line represents the corresponding distribution one gets with ${\cal A} = 2, k = 2$, and a free evolution time of $t = 0.4 \pi/\omega$.}
\end{figure}

\begin{figure}[t] \begin{center}
\centerline{\psfig{height=7cm,file=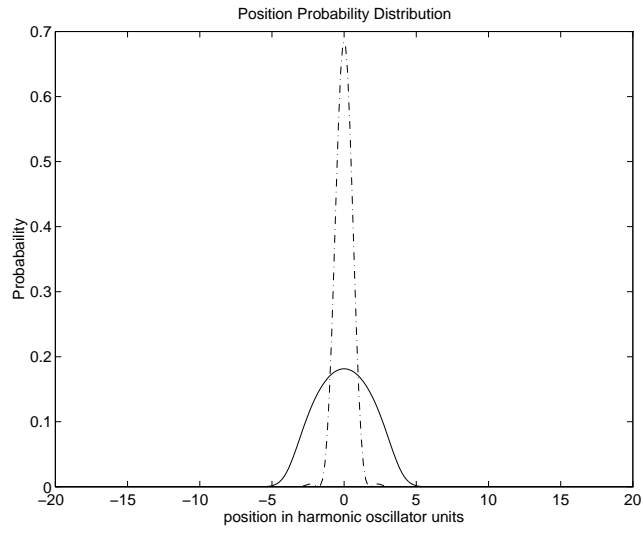}}
\end{center}
\caption{\protect \footnotesize   Solid line: Probability distribution of the trapped condensate. Dot-dash line: Probability distribution when a parabolic  variation on the non linear coefficient is imposed. This is the result at time $t= 0.5\pi/\omega$, where $\omega$ is the trap frequency}
\end{figure}

\begin{figure}[t] \begin{center}
\centerline{\psfig{height=7cm,file=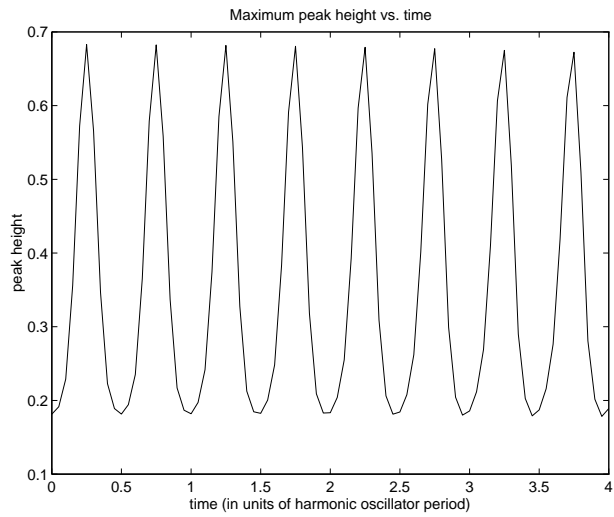}}
\end{center}
\caption{\protect \footnotesize   Variation of the maximum peak height over time in the presence of parabolic $C_{\rm n}$.}
\end{figure}

\begin{figure}[t] \begin{center}
\centerline{\psfig{height=7cm,file=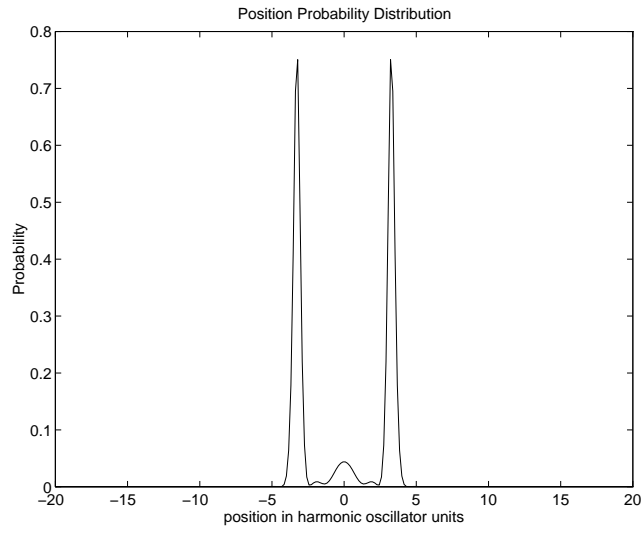}}
\end{center}
\caption{\protect \footnotesize   Probability distribution with a sinusoidal variation of the non-linear coefficient in the presence of the trap.  This is a plot after time $t = 0.4 \pi/\omega$.   }
\end{figure}

\begin{figure}[t] \begin{center}
\centerline{\psfig{height=7cm,file=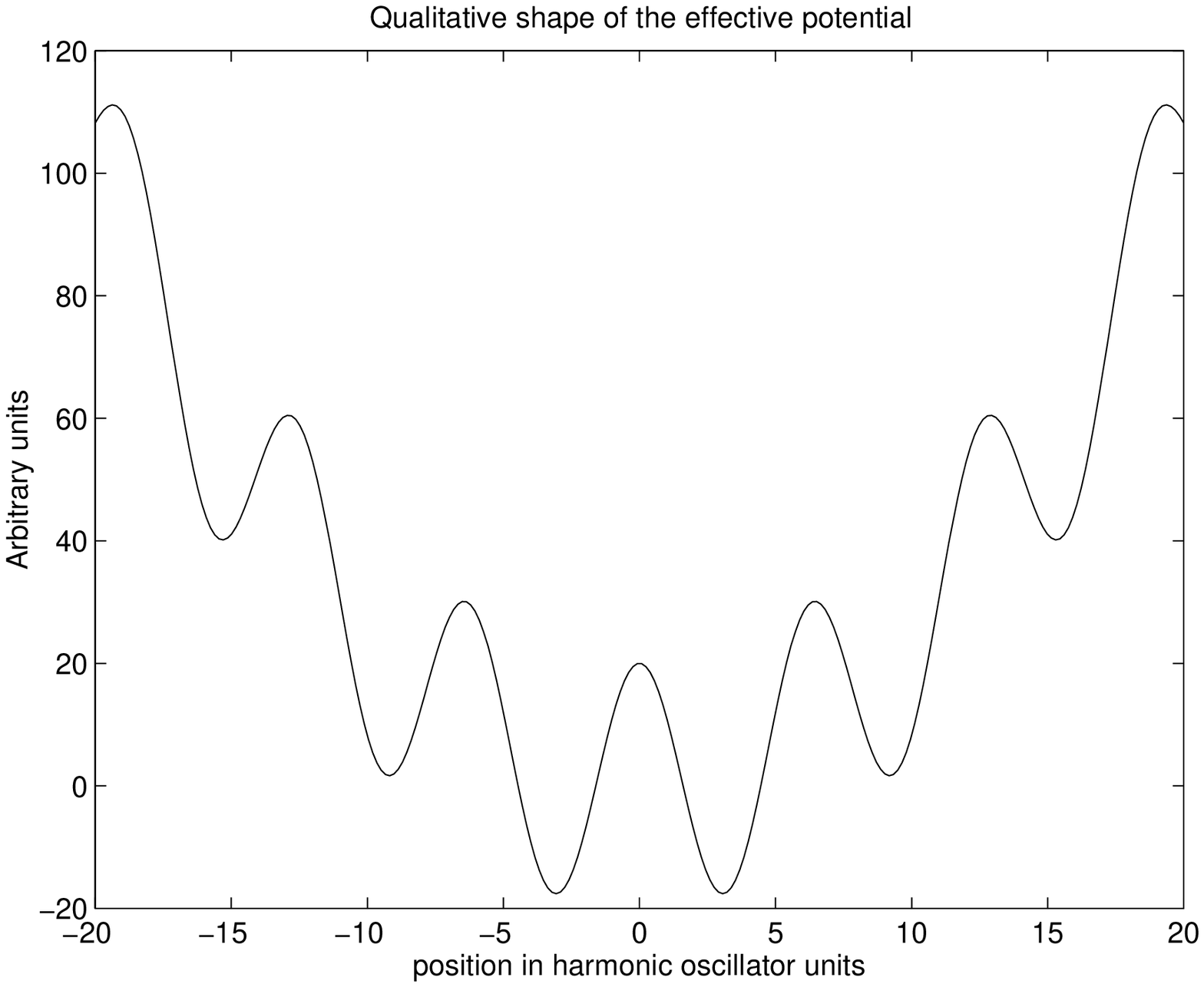}}
\end{center}
\caption{\protect \footnotesize  A qualitative shape of the potential obtained by superposing the two independent potentials on the linear and non-linear terms. This represents an effective potential to generate the probability distribution shown in Fig. 9}
\end{figure}


\vspace{0mm}

\begin{references}

\bibitem{BECRb}  M. H. Anderson, J. R. Enscher, M. R. Matthews, C. E. Wieman and
E. A. Cornell, Science {\bf 269}, 198 (1995)

\bibitem{BECNa}  K. B. Davis, M.-O. Mewes, M. R. Andrews, N. J. van
Druten, D. S. Durfee, D. M. Kurn and W. Ketterle, Phys. Rev. Lett. {\bf 75}, 3969 (1995)

\bibitem{BECLi}  C. C. Bradley, C. A. Sackett, J. J. Tollett, and R. G. Hulet, Phys. Rev. Lett. {\bf 75}, 1687 (1995)

\bibitem{HolBurGar96} M. Holland, K. Burnett, C. Gardiner, J. I. Cirac, and P. Zoller Phys. Rev. A {\bf 54}, R1757 (1996)

\bibitem{WisCol95} H. M. Wiseman, M. J. Collett Phys. Lett. A {\bf 202}, 246 (1995)

\bibitem{MewAndKur97} M.-O. Mewes, M.R. Andrews, D.M. Kurn,  D.S. Durfee, C. G. Townsend, and W. Ketterle Phys. Rev. Lett. {\bf 78}, 582 (1997)

\bibitem{AndTowMie97} M.R. Andrews, C. G. Townsend,  H.-J. Miesner, D.S. Durfee,  D.M. Kurn, and W. Ketterle Science {\bf 275}, 637 (1997)

\bibitem{AdaSigMly94}  C. S. Adams, M. Sigel, and J. Mlynek, Phys. Rep. {\bf 240}, 142 (1994) and references therein.

\bibitem{DroStuSch97}  U. Drodofsky, J. Stuhler, Th. Schulze, M. Drewsen, B. Brezger, T. Pfau, and J. Mlynek To appear in Phys. Rev. A.

\bibitem{LenMeyWri93}  G. Lenz, P. Meystre, and E. M. Wright Phys. Rev. Lett. {\bf 71}, 3271 (1993)

\bibitem{Zha93}  Weiping Zhang  Phys. Lett. A {\bf 176}, 53 (1993)

\bibitem{ZhaWalSan94}  Weiping Zhang, D. F. Walls, and B. C. Sanders Phys. Rev. Lett. {\bf 72}, 60 (1994)

\bibitem{ZhaWal94}  Weiping Zhang and D. F. Walls  Phys. Rev. A {\bf 49}, 3799 (1994)

\bibitem{GPE}V. L. Ginzburg and L. P. Pitaevskii, Zh. Eksp. Teor. Fiz. {\bf 34}, 1240 (1958) [Sov. Phys. JETP {\bf 7}, 858 (1958)]; E. P. Gross, J. Math. Phys. {\bf 4}, 195 (1963)

\bibitem{TieMoeVer92} E. Tiesinga, A. J. Moerdijk, B. J. Verhaar, and H. T. C. Stoof Phys. Rev. A {\bf 46}, R1167 (1992)

\bibitem{SuoHolBur95} Kalle-Antti Suominen, Murray J. Holland, Keith Burnett, Paul Julienne Phys. Rev. A {\bf 51}, 1446 (1995)

\bibitem{NapWeiJul97} Reginaldo Napolitano, John Weiner, Paul Julienne Phys. Rev. A {\bf 55}, 1191 (1997)

\bibitem{EdwRupBur96} M. Edwards, P. A. Ruprecht, K. Burnett, R. J. Dodd and C. W. Clark  Phys. Rev. Lett. {\bf 77}, 1671 (1996)

\bibitem{JinEnsMat96}  D. S. Jin, J. R. Enscher, M. R. Matthews, C. E. Wieman and E. A. Cornell, Phys. Rev. Lett. {\bf 77}, 420 (1996)

\bibitem{Str96} S. Stringari, Phys. Rev. Lett. {\bf 77}, 2360 (1996)

\bibitem{MewAndKur96} M.-O. Mewes, M. R. Andrews, D. M. Kurn,  D. S. Durfee, C. G. Townsend, and W. Ketterle Phys. Rev. Lett. {\bf 77}, 988 (1996)

\bibitem{MewAndvan96} M.-O. Mewes, M. R. Andrews,  N. J. van Druten, D. M. Kurn,  D. S. Durfee, and W. Ketterle Phys. Rev. Lett. {\bf 77}, 416 (1996)

\bibitem{FedKagShl97} P. O. Fedichev, Yu. Kagan, G. V. Shlyapnokov and J. T. M. Walraven  Phys. Rev. Lett. {\bf 77}, 2913 (1996)

\bibitem{HopSav96} J. Hope and C. Savage Phys. Rev. A {\bf 53}, 1697 (1996)

\end{references}
\end{document}